# The Mach-Zehnder Interferometer and Photon Dualism: with an Analysis of Nonlocality

**Paul A. Klevgard**
**Sandia National Laboratory, Ret.**
**pklevgard@gmail.com**

## ABSTRACT

The Mach-Zehnder Interferometer (MZI) is chosen to illustrate the long-standing wave-particle duality problem. Why is which-way (welcher weg) information incompatible with wave interference? How to explain Wheeler's delayed choice experiment? Most crucially, how can the photon divide at the first MZI beam splitter and yet terminate on either arm with its undiminished energy?

The position advanced is that the photon has two identities, one supporting discrete features and the other continuous (wave) features. There is photon kinetic energy that never splits (on half-silvered mirrors) or diffracts (in pinholes or slits). Then there are photon probability waves that do diffract and can reinforce or cancel.

Photon kinetic energy is oscillatory; its cycles require/occupy time. $E = mc^2$ suggests that kinetic energy is physically real as occurrence in time just as rest mass is physically real as existence in space; both are quantized and both occupy/require a dimension for their occurrence or existence. Photon kinetic energy (KE) thus resides in time, but is still present/available for interactions (events) in space; rest mass (e.g., your desk) resides in space but is still present/available for interactions (events) in time. While photon probability waves progress in space and diffract there, photon KE resides in time and never diffracts in space; at reception it always arrives whole and imitates particle impact without being a particle.

Photon probability waves are real; they diffract in space. Acknowledging that the photon has two identities (residing energy and progressing probability), explains photon dual nature. And wave-particle duality is central to quantum mechanics. Understanding it leads to new insights into entanglement, nonlocality and the measurement problem.

What follows, except for Sec. 8-13 (the electron and Schrödinger's Cat), has an alternate, shorter exposition at https://philpapers.org/archive/KLEGSA-2.pdf .

A 30-minute video on nonlocality and photon dualism is at: https://youtu.be/A1Wabkr0YFE

## 1.0 The Photon Requires Dual Identities for Dual Attributes

Our usage of a single word, photon, to refer to quantized radiation leads us to believe there is a single object that has, like any (material) object familiar to us, specific attributes. Unfortunately, in this case the attributes are contradictory: discrete (particle-like) vs. continuous (wave-like). The way out of this impasse is to retain the contradictory attributes but give up on the concept of the photon as an object with but a single identity. For the photon it is imperative to look for a dualism of identities to match up with the dualism of attributes. Toward that end, what follows looks at what is unique about an object or an entity[1] that progresses in only one dimension: the photon progressing in space and the inertial particle progressing (persisting) in time.

When a photon enters the MZI the photon probability of reception gets divided by the first beam splitter such that each arm of the MZI has a 50% chance of photon reception. But if received on one arm or the other, the photon terminates with its undivided kinetic energy (KE). No space device – pinhole, slit, half-silvered mirror – can fractionate the KE of a photon. A photon in a medium (water or glass) has a reduced speed (wavelength), but its energy (frequency/color) are unchanged. Why should this be true? Does it tell us that radiation KE is not well understood?

## 1.1      Rest Mass Invariant over Time; Photon KE Invariant over Space

Rest mass particles and photons are both quantized, measurable entities. They are mirror images of each other in several ways. The photon is stationary in time since anything at the velocity of light suffers infinite time dilation. The force free (inertial) particle is stationary in space within its own inertial system. Being stationary in space makes the particle all rest mass with no kinetic energy (KE). Conversely, the time-stationary photon is all KE with no rest mass. The space stationary particle and the time stationary photon are "pure entities" in that they do not mix KE with rest mass. A "mixed entity" is when rest mass and KE combine to give us familiar matter-in-motion which will be covered presently.

Assume there is a rest mass particle stationary in space. As time passes, successive observers see or measure the same particle. The particle (entity) and its mass remain invariant over successive observations.
Observation invariance over time for material objects is something taken for granted; it is the law of identity viewed temporally. Observers over time share the same, unchanging material object; successive observations have the material object in common. One simply says that time and its passage are orthogonal to existing, space-residing objects. And something is orthogonal to a dimension if it does not reside in that dimension.

- **Inertial (space stationary) rest mass objects reside (occupy an interval) in space and are common to all observations as the object progresses in time.**

For a space-stationary particle, multiple observers differ by their time locations. For time-stationary photons progressing in space, multiple observers differ by their space locations, not by their time locations. As already noted, the single photon traversing the MZI has its probability of reception waves divided between possible observers on either arm. If a photon instead passes through a pinhole or slit, it diffracts into innumerable space paths of probable reception. Each observer on such a space path is a possible recipient of this photon. And each observer would measure the same photon frequency and polarization if termination occurred for them. Photon probability of reception fractionates over space, but photon KE remains invariant. This means the photon has two identities. KE is essential to the photon; KE constitutes the photon's **essential identity**. Potentiality and probability are closely related; probability of reception waves spreading in space constitutes the photon's **potential identity**.

All stationary entities have an essential identity and a progressing, potential-probabilistic identity. The time-stationary photon has its essential identity (KE) while its potential-probabilistic identity due to $E = mc^2$ progresses in space toward termination (reception). A space-stationary carbon-14 atom has an essential identity (rest mass) while its potential-probabilistic identity progresses in time toward termination (decay). All rest mass particle, even the electron, have a theoretical decay point and hence a potential identity.

---

[1] The dictionary definition of "entity" ("something that exists") reflects our preoccupation with material objects (particles). In light of mass and energy equality, "entity" in these pages refers to something involving mass or energy that has a physical presence in a dimension. It may exist or occur and involve space or time.

## 2.0   Photon Essential Identity: Photon KE

The invariance of entities (particles, photons) when progressing through a single dimension (time or space) reflects a common situation: a stationary entity's essential identity residing in one dimension while that entity's potential identity progresses in the alternate dimension. Your desk (or a carbon-14 atom) progresses (persists) in time so it does not occupy an interval there; but it does occupy an interval (volume) in space where it resides. This is how space and time are orthogonal for stationary entities. Stationary entities can only occupy (reside in) an interval/volume in one dimension; moving rest mass entities employ space and time a bit differently as will be outlined presently.

Entities get identified by their essential identity: KE for the photon, rest mass for the particle. This despite the fact that they store something (as a potential identity). So, one can identify a particle as "mass" and write equations for it even though it stores energy. Similarly, one may refer to the photon by its essential identity, namely "photon KE," even though it stores relativistic mass. Equations for the photon ($E = hf$) are written for its energy (for its essential identity).

Since the orthogonal nature of space and time accounts for particle invariance during time progression, it is likely that it does the same for photon KE invariance during space progression. This requires the invariant item to reside in (occupy) but one dimension so that it is unaffected by progression in the alternate, orthogonal dimension. This means that oscillatory photon KE must reside in the time dimension making it common to (shared by) observers on all available space locations (paths) [1, Sec.5].

Stationary entities reside in one dimension and progress (or persist) in the opposite dimension. Residing in one dimension does not prevent them from interaction with the opposite dimension via an event. The rest mass of a carbon-14 atom resides in space but has a termination (decay) event at a time point/location. Photon KE resides in time but has a termination (reception) event (of relativistic mass and its momentum) at a space point/location.

- **Inertial rest mass cannot be assigned a time location because it resides in space.**

- **Photon KE cannot be assigned a space location because it resides in time.**

- **Photon KE in time cannot be fractionated by material, space-residing devices: pinholes, slits or half-silvered mirrors.**

- **By residing in time, photon KE is orthogonal to photon space paths making this energy common to (shared by) all possible space observers.**

A quantized, existing particle entails rest mass which requires (occupies) a space volume. A quantized, occurring photon entails oscillatory-cyclical KE which requires (occupies) a time interval. Photon KE cannot be a mere quantity; it must involve oscillation cycles occupying time. The concept of photon KE as a mere quantity without oscillation and with no presence in a dimension is wrong.

Photon energy is created when work is done upon a charge. Photon energy can also be generated by the release of stored work (an electron changing atomic orbits). Doing or releasing work to produce radiation creates photon KE residing in time as pure oscillation rather than the oscillation of something material or existing. If you ask how pure oscillation can exist, you are betraying your bias for a material reality that only *exists* (and in space). Photon energy oscillation **occurs** and it does so in time. Occurrence in and of itself is the ontological counterpart of existence in and of itself; the former is time-residing energy, the latter is space-residing mass. The realm of energy/occurrence should be granted equal standing with the realm of mass/existence just as $E = mc^2$ implies.

- **Photon KE is matter-free oscillation residing in time.**

- **It constitutes the photon's essential identity.**

---

Rest mass is an entity's essential identity occupying space. Letting photon KE be an entity's essential identity occupying time presents conceptual challenges: 1) **broadening our current concept of what is real**; 2) **envisioning photon KE as pure oscillation;** and 3) **being common for a dimension.**

**Reality**: We all live in a world of material objects that occupy space. Our concept of reality is rooted in existence, mass and space. But radiation is based on occurrence, energy and time; trying to explain it based on our existing material world only leads to paradoxes.

Entities require a presence in a dimension. Most regard the photon as an entity, but then try to place its KE in space as a quantitative payload of a real particle. Regarding photon KE as a mere quantity traveling in space is an adaptation of a 19th century concept characterizing matter-in-motion. Relativity and quantum mechanics made equality foundational: space with time, and mass with energy ($E = mc^2$). Quantized mass is an entity, but so is quantized radiation; they are mirror existence/occurrence images of each other. Quantized energy entities in time are the ontological counterparts of quantized mass entities in space.

**Pure oscillation**: Physicists have embraced the oscillation of nothing or the oscillation from nothing: vacuum state fluctuation is an essential part of QFT. But this latter oscillation resides in space (of course), appears randomly and creates (virtual, transient) particles that cannot be measured directly. This is unlike the oscillation of photon KE which can be measured, doesn't depend upon particles, hypothetical or real, and whose origin is real work.

The notion of an immaterial, oscillatory photon energy occurrence residing in time is no more problematic than an un-measurable, transient, harmonic/oscillatory, virtual particle residing in space. And it completes the symmetry of particle mass-in-space with photon energy-in-time.

**Being common**: That an existing rest mass object in space is common for (shared by) observers in orthogonal time is a concept familiar to us. Occurring photon KE being common for observers in space is the equivalent, but it is an unfamiliar concept for us. The importance of that concept will appear presently.

---

### 2.1 Photon KE Mimics Particle Impact

The photon's essential identity operates (oscillates) in time. It is the photon's potential identity (potential mass, next section) that operates (progresses) in space. Because photon KE is common to space paths it is available for probabilistic release events on those paths its waveform potential identity traverses. But being in time also places some limitations on how photon KE can interact with matter.

Since the KE of a photon resides in time while the rest mass of a target resides in space, they are orthogonal to each other. With one occupying time and the other occupying space, the only way they can intersect is at a joint time-and-space point, namely an <u>event</u> since the latter requires the participation of both rest mass and KE. Hence time-residing photon KE can release to (intersect) an orthogonal dimension (space) only as a discrete event; that is, at a space point thereby mimicking particle impact. This energy transfer is the discrete/particle aspect of photon behavior. It is also random on an individual basis, something that bothered Einstein whose preference was always for strict causality. All of this is a direct consequence of time-residing, quantized photon KE being forced to access space-residing matter via discrete events. Of course, everyone wants to interpret photon KE reception as particle impact; but this is to impose our common, material world experience onto the realm of radiation where it does not apply.

- **Because photon KE resides in (occupies) time while rest mass resides in (occupies) space, the only way they can intersect is via a reception event that is discrete in both dimensions.**

- **Such an event is invariably interpreted as particle impact to conform to our concept of reality as limited to existence, mass and space.**

## 3.0    Photon Potential Identity: Mass Stored via $E = mc^2$

Our first photon identity, kinetic energy (KE), accounts for a number of photon attributes. These include: 1) oscillation; 2) non-rarefying energy available on diverging space paths; and 3) quantization, i.e., occurrence (a cycle) is whole just as existence (a particle) is whole.

This leaves a number of attributes for our potential identity to contribute: 1) probability of photon reception; 2) spreading and progressing on all available space paths; and 3) collapse of what fills those space paths.

The probabilistic nature of photon reception points to something latent, yet ready to emerge. Something that facilitates the energy transfer, but whose space presence depends upon photon KE itself. Such an intimate dependency must be that of $E = mc^2$ storage. One may conclude that the photon's second identity is its potential (stored) mass.[2]

- **The photon has two identities: essential, residing in time, and potential, progressing in space.**

A photon's essential identity is oscillatory energy which is kinetic (unstored), operates (oscillates and resides) in time making it common to space paths. Its alternate, stored, orthogonal identity operates (progresses) in space by filling out available space paths. Because photon KE oscillates, so does its potential mass. Since the latter progresses in space while oscillating, it has the (continuous) waveform.

---

[2] Potential (aka, relativistic) mass is out of favor these days with many physicists, largely for pedagogical reasons ("don't confuse students!"). Some wish to replace potential mass with energy arguing that the latter sustains potential mass and therefore potential mass is the same as (kinetic) energy. This argument is not convincing. Stored thermal energy sustains a mass increment in the body that hosts it, yet no one says that thermal energy is the same as mass. If you wish to deprecate potential mass, then you should also deprecate potential energy; they both have something physical (mass or energy) being stored. In these pages <u>stored mass</u> or <u>potential mass</u> will be used for what KE stores. Potential mass is what the photon has: it can be measured; one should be loath to deprecate what can be measured [2].

When visible light enters glass or water its time-residing energy is unaffected; hence the light's frequency and color are unchanged. But velocity is diminished because wavelength is shortened; this means momentum p increases according to $p = h/\lambda$. Photon momentum is a consequence of photon potential (relativistic) mass. Momentum change in a medium confirms what has already been put forth: the photon's space-progressing identity is a waveform of momentum-bearing potential mass.

It was noted (preceding section) that the photon's essential identity yields the photon's particle-like nature, namely termination at a point. It is the photon's potential identity that yields the "continuous" aspect of photon behavior permitting wave interference.[3]

- **Photon KE entity is pure occurrence in time; particle rest mass entity is pure existence in space; each is the essential identity by which we know the entity; each has a potential (stored) identity.**

- **Photon KE occurs and resides in time making it common to those space paths its potential identity traverses.**

- **Photon potential (stored) mass also occurs and progresses and rarefies as a waveform on multiple space paths. Its local intensity determines probability of photon KE reception.**

Physicists deny photon potential mass a space presence; to them it is merely a quantity explaining photon momentum. Photon potential mass and photon KE have suffered the same fate at the hands of physicists; both are regarded as mere quantities without a presence in a dimension. This view is wrong; it dates from the 19th century and consequently denies the equality of energy with mass. You can't reject dimensional presence for the photon's essential identity (its KE) and then argue that the photon is physically real.

Photon potential mass progresses in space at the speed of light while sharing in the oscillation of its opposite (energy) number; this space-presence of something stored, plus velocity and oscillation create the probability wave character of the photon.[4] This waveform's space presence is real, but in an occurring, potential, hence probabilistic way. It is continuous in space and can disperse and rarefy there; but via interference, wave crests can superpose and reinforce. With potential mass rarefying in space, its momentum follows suit. But at photon termination, potential momentum collapses and reverts to its classical form: a quantity and a vector with the latter pointing from source to target.

Our physical instruments cannot capture or measure this wave directly; they only receive photon KE or momentum. Nonetheless, from experiments one can infer two of the unusual properties of photon potential mass. It is: 1) a wave of "objective probability [3, p.47-8]"; and 2) capable of instantaneous collapse.

**3.1    Objective Probability Waves**

The diffraction pattern of coherent photons passing through a pinhole (the representation is a so-called Airy pattern) can be predicted from a relatively few parameters. The mathematics yields areas of high and low wave intensity on a target screen. No one doubts that the mathematics is modeling something real. The mistake is to regard it as modeling the photon as a unitary object; rather it is only modeling the waveform space-progressing identity of the photon, namely photon potential mass. The local intensity of this wave determines the probability of where photon KE will terminate on a space-residing target. But time-residing photon KE and a space-residing material target are dimensionally orthogonal; they can only interact via a discrete (reception) event that is random and noncausal for the instance but is predictable for the aggregate based on the potential mass waveform.

---

[3] Commentators place the photon's wave nature and particle nature on equal footing. They fail to notice that particle-like behavior depends on KE transfer but wave behavior depends on the potential for reception (probability). The two are kinetic vs. potential and related by $E = mc^2$.

[4] There is, of course, an EM wave character as well, created by work done on a charge: orthogonal, self-sustaining electric and magnetic fields that are in synch with oscillatory photon KE. These EM radiation cycles as pure occurrence (no mass) are present for space paths and can interact there with matter.

**3.2  Collapse of Probability Waves**

A photon's potential mass progressing on all available space paths is an $E = mc^2$ expression of that photon's KE. This makes the entire wavefront of potential mass dependent upon a single oscillation entity in the time dimension. If that occurrence in time ceases, the dispersed potential mass in space disappears (ceases to occur) in its entirety. And the latter collapses instantaneously regardless of its spatial extent with no signaling required.

Instantaneous collapse happens because: 1) occurring potential mass carries neither energy nor rest mass; and 2) occurrence (i.e., oscillation frequency) or cessation of that occurrence in the time dimension is common to all space paths. Having something immaterial in space depend upon pure oscillation occurrence in time explains instantaneous collapse. This is another strong indicator that photon energy resides in time rendering its oscillation common to all space paths.

<u>Local collapse without termination</u>:    Assume some of the available space paths for photon potential mass waves are blocked by a detector. If the photon KE does not register (terminate) on that detector, then those blocked waves will collapse without a trace. Waves that cannot progress in space cease to occur. (If rest mass particles cannot progress in time, they cease to exist.)

<u>General collapse due to termination</u>:    At photon termination (reception) all remaining photon potential (stored) mass waves will collapse regardless of how widely dispersed they are. The result is the delivery of the photon's undiminished KE (and momentum) to a space point on the target.

Assume for the moment that our Sun only emits a single photon. We like to imagine this photon as travelling through space as a packet/particle that reaches us in 8 minutes. Upon reflection we realize that this single photon has a probability wavefront that controls termination. This wavefront expands (and attenuates) in space along innumerable possible termination paths. Our tiny earth can only block a small segment of this single photon wavefront and what is blocked is likely to collapse without triggering photon KE termination. The remaining, unblocked wavefront continues into deep space to occasion termination on another planet or star; but more likely never to terminate, with oscillatory photon KE stuck in the time dimension, while still common for the space dimension and, sans termination, producing no illumination (is this dark energy?).

**4.0  Photon as Particle?**

Regarding the photon as a real, path-traveling discrete particle runs into various paradoxes. Nevertheless, it is still a popular concept for physicists. Classical physics has KE as a quantitative payload for something moving in space and quantum physics has never challenged that concept for massless radiation. Of course, there is the comparison with the electron, based on their shared wave behavior. But to lump the photon and the electron together as examples of wave-particle dualism is too simplistic. The photon has only waveform; its only presumed particle nature is termination at a point and that is misinterpreted. The electron truly has both wave and (rest mass) particle features. The electron leaves a trace in a cloud chamber because it follows a trajectory; the photon does neither.

Physicists get away with calling the photon a particle because the photon's quantized energy acts like a particle when interacting with their instruments. Abraham Pais [4, p.350-1] writes that although the photon has zero mass, physicists "… nevertheless call a photon a particle because, just like massive particles, it obeys the laws of conservation of energy and momentum in collisions, with an electron say (Compton effect)."

Physicists want to write equations that describe the transmission of energy or force over space; that is the basis of their craft. Waves are not suitable for that since they disperse; hence "particle" is the concept of choice to traverse space. It also conforms to the universal belief in existence, mass and space as defining reality. Physicists have their computational reasons for regarding the photon as a particle, but that doesn't make it a real particle.

## 5.0	How the Photon Traverses the MZI

A single photon entering an interferometer's first beam splitter (BS) has its space-progressing potential mass divided in two while its time-residing KE is unaffected. If the upper path of an interferometer contains a photon detector (obstacle), the photon's (reduced) potential mass wave front will reach it but with only a 50% chance of terminating on it. If photon reception (termination) **does not** take place on this (blocking) detector, then local collapse of these blocked waves occurs (Section 3.2). This means the lower path instantaneously (nonlocally) converts from 50% to 100% probability of photon reception since the competing path has been eliminated. Stored (potential) mass progressing on space paths and constituting an immaterial, waveform, probability occurrence can be retracted instantly if it cannot oscillate and hence cannot progress on that path due to an obstacle. Stored (potential) mass is a collapsible "ghost wave" of probable release.[5]

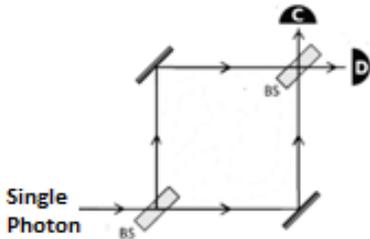

A single photon allowed to traverse both arms (no blockage) of a Mach–Zehnder interferometer yields wave information (interference) when it encounters the second beam splitter. This is because the single photon's potential mass – divided by the first beam splitter – undergoes interference when reunited by the second beam splitter. If you place a detector on one arm of the interferometer to obtain "which-way" (which path) information, you block the passage of the photon's stored (potential) mass on that arm. If the photon does not register (terminate) on that detector it is a mistake to conclude that nothing was on that path and that the photon chose the other path. If the photon does register on that detector, it is also a mistake to conclude that nothing traversed the other path.

Wave interference occurs in space; it is the space-progressing potential mass of the photon that produces this. Blocking the passage of photon's potential mass wave on one arm prevents any wave interference at the second beam splitter. Similarly, for a photon traversing a double slit, positioning a detector behind one slit has the same effect as a detector on one arm of an interferometer; the wave pattern disappears. Space location ("which-way") and wave interference phenomena (multiple paths) are mutually exclusive. The blocking of probability wave paths constitutes a physical change for radiation even if photon reception does not occur on the obstacle/receptor.[6]

It was, and remains, a great mystery that a photon seems to adjust its behavior – interference or no interference – when an experimenter places or removes her measuring (blocking) device on one path/arm even when that device fails to receive a photon (i.e., a termination). This has occasioned many "which-way" experiments over the decades. Certainly, the most famous was suggested by John Archibald Wheeler. His delayed choice thought experiment [7] has generated a huge literature and several attempts to carry it out in practice.

Wheeler theorized that the photon made a defined choice at the beam splitter: follow both paths as wave, or follow a single path as particle. Subsequent to that choice the observer might insert (activate) detectors on the two paths to measure (receive) a particle, or, retract (deactivate) those detectors and measure a wave (interference at the second beam splitter). This role for the observer in determining wave versus particle aligns Wheeler with his mentor, Bohr, who argued that reality depends upon how one decides to measure it, a view anathema to Einstein. For Wheeler the role the observer plays implies retrocausality: the first beam splitter's choice gets determined by the subsequent observation choice. This led Wheeler to claim that "we…have an inescapable, an irretrievable, an unavoidable influence on what we have the right to say about what we call the past."[7]

---

[5] The term "ghost wave" is from Einstein [5, p.2-3]. His instincts, as usual, were correct about a retractable wave of probability. But the irony is that Einstein tried to eliminate relativistic mass; perhaps his biggest oversight since he first equated mass and energy.

[6] Such a physical change is the basis of interaction-free measurement. See Elitzur–Vaidman bomb tester [6].

Wheeler was wrong; there is no such thing as retrocausality. Nor does a photon object make a "choice" at the first beam splitter. When particle detectors are placed on both paths only one detector will receive the singular photon KE. This leads to the too-easy assumption that nothing travelled the other path. In fact, the photon's potential mass travels both paths. Once again, the mistake here is to limit the photon to but a single identity and overlook its probability identity (potential mass). The in-flight photon is a pure, waveform probability occurrence dependent upon time-residing energy that doesn't follow space paths. The naïve idea that anything "real" will register on our material detectors on a known path discounts probability waves. These waves are physically real, occur and make their own arbitrary choice as to whether to register or not and if not, then to collapse without a trace.

## 6.0    The Photon Summarized

Successive generations of physicists have used the MZI (and the double slit) to investigate duality and the nature of radiation. It cannot be said that their efforts have advanced our understanding much. The photon still gets regarded as a unitary object – either as a quasi-particle or as a field disturbance – that has contradictory attributes.

- **The two identities explain all the usual photon issues: 1) how the photon can split at the MZI's first half-silvered mirror yet keep its energy undivided; 2) why an obstacle on one MZI path destroys wave (interference) behavior even when the photon does not terminate on that obstacle; and 3) why photon termination of diffracting coherent light is deterministic in aggregate but random for the instance.**

- **Radiation has been interpreted with the concepts used for matter (particle physics); this requires energy to be a quantity with no presence in a dimension. This mistake leads to numerous paradoxes.**

### 6.1    Constant Speed of Light

If the photon was a packet of energy moving through space its velocity would differ for different inertial systems. The same is true if the photon was a wave disturbance of a "stationary" medium, the aether. Both of these interpretations have proven to be untenable; they are the result of us applying massy-body mechanics to radiation.

With a photon energy residing in time, what remains to function in space are two immaterial, occurring, rarefying waveforms: EM waves and potential mass waves. Both will collapse instantaneously regardless of extent since as occurrences they depend upon photon energy in time. EM waves permit our airborne communications; potential mass waves govern probable photon termination on a target.

---

[7] Wheeler's oddly-worded statement [8, p.6] stops just short of asserting that one can change the past. His mentor did not excel at clarity either! For views of Wheeler, Bohr and Einstein see [9].

The potential mass constitutes a wavefront of probable reception that progresses at the speed of light without carrying photon energy with it. Both waves have a constant phase velocity of wavelength divided by cycle time. If an observer moves toward (away from) the photon source, she will diminish (increase) both the wavelength and the cycle time. The phase/wave velocity stays the same although the wavelength-momentum changes, as does the energy-frequency and hence the color for visible light. Energy variation between observers reflects the work done by, and velocity of, each inertial system relative to the photon source.

Einstein's second postulate – the speed of light as constant – was a positive contribution in 1905 when so little was known about radiation. But putting photon energy in time allows one to recognize that "photon velocity" is simply phase velocity; this makes photon constant velocity a straightforward wave feature. And a postulate is not needed to enunciate an explicable feature.

### 6.2   Conceptual Obstacles

Photon physics leads to unresolved paradoxes because we approach it with the wrong (particle-centric) ontology; that is the first obstacle. The difficult concepts in these pages include orthogonal identities, KE residing in time, nonlocal collapse, probability waves, and occurrence-energy-time as the equal of existence-mass-space. But without these concepts there is no explanation as to why the photon rarefies in space yet keeps its energy intact.

Another obstacle is our very human tendency to apply familiar constructs and objects to the realm of radiation. We are heir to a 19$^{th}$ century concept of KE as a formless quantity with no dimensional presence and no oscillatory character; it does not serve us well in the case of the photon. Radiation is the transmission of KE over space and a different concept of KE is required there, one that incorporates oscillation. But it is so easy and comfortable for us to think of reality as limited to existence, mass and space with KE as quantity forced to fit in as best it can.

Retaining the traditional view of KE as quantity and continuing to apply material world concepts to radiation leaves us unable to explain something as simple as the MZI. In lieu of an explanation we resort to makeshift explanations such as complementarity. Old and familiar ideas are comfortable; change is difficult; that is a final, big obstacle. As Abraham Pais writes, "[L]ike most of humanity, physicists tend to cling tenaciously to what they know or think they know, and give up traditional thinking only under extreme duress [10, p.137]."

Experiments may never be able to confirm that photon KE resides in time; our instruments exist in space and are limited to quantitative measures of events, not entities per se. Nevertheless, entanglement is strong, indirect evidence that photon KE resides in time.

### 7.0   Photon Entanglement in Time

Entities should have similar bonding abilities whether they exist or occur. If particles or atoms can bond together in space, then photons should be able to bond together in time. When entities bond, they become parts of a common object, something familiar to us for material objects; thus, a sodium ion bonds to a chloride ion to give us salt. These two ions can bond only if their essential identities (their masses) are adjacent in space. Photons can bond only if their essential identities (their energies) are adjacent in time. They achieve this if they are the product of a common event. This can happen if a photon interacts in a way to divide into two new photons; the latter are then bonded (entangled). Bonding in space accords with our existence-mass-space view of reality; bonding in time is equivalent, but with different entities and dimensions; it is an alien concept for us.

Suppose a high energy photon enters a crystal and divides into two lesser, entangled photons, one blue and one green in frequency. The "parent" photon's kinetic energy is in time; the kinetic energies of the two daughter photons are also in time and are adjacent there. These blue and green photons remain distinct; hence they retain their frequencies and they can terminate independently. They also send out their potential mass wavefronts in all directions; these waves of probability are subject to instantaneous collapse.

Mass-based entities (particles) and energy-based entities (photons) bond to their like in the one dimension where they reside and occupy an interval, space and time respectively. This makes them common to (shared by) observers in the alternate dimension. Space-entangled atoms or molecules are common to (shared by) time-separated observers. Such observers all encounter the same entangled pair despite their time separation.

- **For space-entangled particles progressing (persisting) over time, their essential identities (rest masses) are shared by multiple time observers. Your desk is shared over time by all your family members who use it.**

For entangled photons, their probability of reception waves spread out as wavefronts. At any one point in time, numerous detectors (observers) sharing the leading edge of an expanding wavefront have the possibility of photon energy reception.

- **For time-entangled photons progressing over space, their essential identities (kinetic energies) are shared by multiple space observers.**

On creation, our blue and green photon KE entities reside together in time and their spin orientation as a unit is zero. Meanwhile their probability of reception wavefronts progress in space at the speed of light. At some distant point the blue photon's wave may trigger a blue photon reception and several things then happen.

The blue photon's KE is transferred from time to a rest mass space target via a space/time event (absorption). The blue photon's potential mass (probability) waves in space collapse instantly, nonlocally, regardless of extent. In addition, the spin of the blue photon is defined which simultaneously defines the spin of its time-conjoined partner. No space signal is required to orient green photon spin since the two photons' essential (KE) identities are not even in space.

Photon KE is never in space in the sense of occupying a volume there and having a defined location; it is merely common to, and therefore available for, all space observers by virtue of being in orthogonal time. This deceptively simple concept is actually very difficult for us because we only think in terms of the classical reality framework (ontology) of every entity (photon included) having a defined location in space at a time point. But stationary, pure entities (mass without KE or KE without mass) can only have a location in the single dimension where they reside, and photon KE does not reside in space.

So, the terminations (receptions) of paired photons individually at widely separated space locations does not mean that the two photon essential identities (KE), are space separated. Consider the opposite, if you have two decay-prone carbon-14 atoms bonded (entangled) in space and they terminate (decay) at widely separated time locations, you are not going to say that the two carbon-14 essential identities (rest masses) are time separated. That is to project the time location of a termination (decay) event back onto an entity (carbon-14 atom) that residing in space never had a defined time location. But everyone does exactly that when applying the space location of a photon's termination (reception) event back onto the photon as a KE entity.

Entangled photon energies are time-residing occurrences common to (available for) all space locations. The puzzle of nonlocal interaction is created by interpreting the photon within our particle-centric reality where all entities must have a location in space, as opposed to occurring entities being simply common for space.

Bottom line: we assume that the two (distant) photon termination events mark photon object locations in space and conclude that their spin coordination must be in space. Hence, we interpret photon nonlocality via an incorrect (particle) concept of the photon itself.

- **Photon entanglement and its supposed nonlocal change is the best proof that photon KE is an entity residing in the time dimension and able to bond there.**

- **The essence of a photon, its KE, resides in time and can bind there to another photon's KE. The reception of one photon at one space location defines spin for both time-adjacent, photon essential identities; this without any signaling or coordination across space.**

So, Einstein is right and his critics wrong. There is no "spooky action at a distance" because with photon KE in time there is no distance. John Stewart Bell would be very pleased with that.

Entanglement is the bonding of like entities: rest mass bonded to rest mass; photon KE bonded to photon KE. But Nature is clever and subtle; she also allows for the union of dissimilar entities – rest mass with KE – giving us mixed entities (matter-in-motion) that: 1) reside in both dimensions, space and time; and 2) possess both forms, particle-form and waveform. The idea that matter-in-motion might have a wave character was first put forth by Louis de Broglie in 1924.

## 8.0   De Broglie's Wave Theory

Louis de Broglie was more of a speculative philosopher than a physicist. He believed that light quanta had (very tiny) rest mass which was not constant and that particles could be regarded as thermodynamic machines [11, p,1054]. He took one incorrect assumption – that waveform light had a (rest mass) particle nature – and argued the reverse, namely that rest mass particles must have a waveform nature. He equated particle (persisting) rest mass where time has no relevance to a radiation wave of energy $hf$ where (cycle) time is all-important; this did not make much sense.[8] De Broglie's 1924 thesis constitutes a "…barrage of novel ideas and confusing developments … [11, p.1047]." Textbooks and historians of physics rightly laud de Broglie for opening the way to a true quantum theory. But those few [12] who examine his thesis closely agree that his (shifting) arguments do not support his conclusion; he achieved spectacular success based on wrong supporting ideas.

Physicists have never paid much attention to de Broglie's theoretical arguments; their focus has been on the detection of electron waves (Davisson and Germer) and then on the Schrödinger wave equation. As a result, there is currently no convincing explanation for matter waves. But the reason that both EM radiation and the moving particle (electron) have wave characteristics is much simpler than de Broglie imagined. Both the photon and the moving particle involve work done and the resulting KE is always oscillatory.

## 9.0   Matter-In-Motion as a Mixed Entity

The traditional, 19th century view that the quantum pioneers inherited is that KE is a mere quantity possessed by matter-in-motion. This concept does not apply to the photon; it turns out it doesn't apply to matter-in-motion either. There are not two types of KE, one for matter-in-motion and one for radiation; kinetic energy is always an entity occurring (oscillating) in time just as rest mass is always an entity existing in space. Stored energy doing work upon rest mass creates KE as an oscillation entity in time while simultaneously joining it with rest mass entity in space.

---

[8] $E = mc^2$ is a conversion equation. If one side (E or m) is kinetic (unstored) then the other side (m or E) is potential (stored). Thus unstored (rest) mass has a great deal of stored (potential) energy. De Broglie did not observe this distinction and equated rest mass, unstored, with radiation KE, also unstored. He did not recognize that the particle and the photon had inverse ontologies: existence/mass vs. occurrence/energy. Photon ideas were so new in the early/mid-1920s!

So, matter-in-motion is **not** what we have been taught, namely the adding of KE-as-**quantity** to rest mass. Rather, matter-in-motion is the union of two separate entities: existing, non-oscillating rest mass entity plus occurring-oscillating KE entity. Matter-in-motion is a mixed entity. Pure entities (inertial rest mass, photon) have a location and reside in just one dimension; an inertial mass, stationary and residing in space, has location relative to a reference mass; a photon, stationary and residing in time, has location relative to a reference photon.

Our mistake – fostered by mechanics and our use of Cartesian space-time graphing – has been to assume that all entities have defined location in both space and time. But even when moving in space a rest mass object does not have a defined location in time. Only an object's putative perception/measurement **event** has a location (defined position) in time. That is, it is the perception/measurement event (real or imagined) that has a defined (real or imagined) time location, not the rest mass itself. In the case of the photon, by failing to distinguish between object (photon KE entity) and event (KE reception), we assign a space termination location to the photon when its essential identity does not even reside in space (until it becomes an event).

Mixed entity behavior is most evident for electrons. They are never quiescent so they always have significant KE relative to their mass and a resulting de Broglie wave character in space. High speed electrons have more KE relative to their rest mass, but their wave character becomes harder to measure. This is because cycle wavelength and particle momentum are inversely related; relatively low speed, low momentum electrons have longer wavelengths making diffraction detection easier. Davisson and Germer were the first to detect matter waves in 1927 using relatively slow electrons.

## 9.1  Waves Affect Rest Mass via Momentum

When work is done on an electron, both the electron rest mass and its charge get their share of the KE that work creates. Acceleration of electron charge creates EM radiation. Acceleration of electron rest mass entity adds KE entity to that rest mass creating a mixed entity. This mixed entity has its rest mass in space; also in space is the potential mass of the KE created by accelerating the electron rest mass. It is this space-progressing potential mass waveform (or wave packet) that interacts with the electron's small rest mass (also progressing in space) and does so via shared momentum.

Potential mass is the $E = mc^2$ expression of the KE entity joined to electron rest mass entity. Potential mass has momentum; recall that the photon has momentum due to it potential mass. Electron rest mass also has momentum since it always has velocity/movement. Momentum is a quantity dependent upon both mass and velocity; it is also a vector with a space orientation. The momentum of waveform potential (stored) mass has a constantly changing vector orientation; this fluctuating directionality is impressed upon (merged with) electron rest mass momentum.  As a result, the electron's rest mass will imperfectly take its momentum direction from waveform potential mass; and the smaller the rest mass the more perfect is the momentum coordination and the resulting wave motion of the rest mass. The wave function $\Psi$ allows us to model the electron's potential mass waveform; from this one can predict electron rest mass space paths/locations in the aggregate.

## 10.0  Particle Energy Entanglement

Kinetic energy resides in time and this is the case if it has joined with rest mass (creating a mixed entity) or if it is not so joined (a pure entity). Whether pure or mixed, instances of KE can bond (entangle). Section 7.0 covered the entanglement of pure entities (photons). Such energies are time-adjacent because they are the product of paired photon creation: a high energy photon split within a crystal; or an electrically excited semi-conductor (quantum dot) producing photon pairs. However, the energy entanglement of mixed entities typically involves energy sharing, not pair creation.

Energy sharing/combining is the case with valence electrons in a conductor at low temperatures. The energies (oscillations) of these electrons combine in the time dimension with lattice energies (phonons) to create Cooper pairs [13, p.86-89]. The separate oscillations of each electron have now merged into a single occurrence in time. Two Cooper-paired electrons are spin entangled much like two entangled photons. Because the occurrence bond is in time, the two electron rest masses may be separated (carefully) in space without breaking their entanglement [14]. This is further proof that KE resides in time, not space.

Electrons emit a photon when they are laser (photon) pulsed. If an electron has its spin manipulated to be undefined (spin superposition), then when laser pulsed, the emitted photon: 1) has its spin polarization in superposition; and 2) has its energy entangled (in time) with the electron's energy. If two such electrons are made to emit entangled photons, then an optical combining of the two photons can in turn entangle the energies of the two electrons [15]. Researchers find ever more ingenious ways to exploit these possibilities.

There is a general understanding that the entanglement of separated particles depends upon energy and coherent/shared oscillation. But no one thinks of the electron as a union of rest mass entity in space with KE entity in time. Since particle entanglement is actually KE entanglement and in time, the (presumed) nonlocal changes are confined to energy related features, principally spin/polarization and angular momentum. Static rest mass related features do not undergo "nonlocal" change.

There is a type of non-energy bonding ("clumping") that occurs when gas atoms are cooled close to absolute zero and become a Bose-Einstein condensate. These atoms have almost entirely lost their KE entities; with such low energy and momentum, their wavelength becomes large (ca. a micron); they almost cease to be mixed entities. They are pure, indistinguishable matter quanta existing and residing in space with some very strange properties.

## 11.0     The Double Slit for Photon and Electron

The photon has no rest mass; it passes through the double slit simply as a wave of potential mass. These waves interfere and their local intensity determines probable transfer of time-residing KE to a target at a space point.

When an electron enters a double slit its rest mass must pass through one slit or the other. But the electron's de Broglie waves of potential mass will pass through both slits and interfere. The momentum sharing between waveform potential mass and electron rest mass has the rest mass tending to follow waveform intensity as just outlined (Section 9.1).

### 11.1     Duality, uncertainty, indistinguishability and collapse

**Duality:**     Radiation has two immaterial oscillations moving (expanding/rarefying) at the speed of light in space: a waveform of self-sustaining electric-magnetic fields synched with a waveform of potential (relativistic) mass. The former can excite free electrons in an antenna; the latter determines local probability of photon termination upon a target. As we have seen, this termination of time-residing photon KE at a point gets incorrectly interpreted as the arrival event of a discrete particle. With no particle involved, there is no wave-particle duality for radiation. In contrast to the pure entity photon, the moving electron as mixed entity is both a wave (of potential mass) and a particle (of rest mass).

- **Entities have two identities which function in two dimensions. But our physics overlooks this and confines reality to what can be perceived or measured in space: rest mass; EM waves; and received KE and momentum both regarded quantitatively (no dimensional presence). These constituents get lumped together within a particle-physics framework with no distinction drawn between essential (unstored) constituents versus potential (stored) constituents. Little wonder that duality gets misinterpreted.**

- **The photon is not dualistic; the moving electron is.**

**Uncertainty**: Uncertainty is a consequence of a quantal rest mass particle having a wave character. There is a tendency to regard wave behavior as a brute fact of matter. In reality, wave behavior is a natural consequence of KE being oscillatory. Uncertainty only characterizes rest mass if the latter possesses KE; only mixed entities have position-momentum uncertainty.

The actual composition of a matter wave is not understood since potential mass is overlooked or deprecated by physicists. When joined with rest mass, time-residing oscillatory KE has no space presence save for its alternate, stored identity, namely waveform potential (relativistic) mass.

Uncertainty results from blending existence/mass with occurrence/energy at the quantal level where the two are comparable in effects. It does not apply to pure entities so it is not universal, although it has been so construed.

**Indistinguishability:** Quantum theorists devote a lot if ink and mathematics to quantum particles being indistinguishable. This "feature" of the quantum world has a straightforward ontological explanation. Rest mass entities (particles) occupy space and have a location there. They can be distinguished by their location since no two rest masses can occupy the same space. But electrons don't conform well to this requirement. Locating them in space is beyond our capabilities. We characterize electrons not with coordinates but with a wave equation; and for a typical atom we portray electrons as overlapping clouds occupying a space region.

Our inability to define the space location of an electron is not simply a matter of it being tiny; nor of its erratic movements. The real issue is that the electron's time-residing KE dominates what mass there is. Kinetic energy is not even in space so the more it dominates a particle the more "ill-defined" will be that particle's space location; the endpoint of this is the photon whose essential identity (energy) has no location in space. The smearing of electron space location is ontic, not epistemic.

Rest mass in space and KE in time always seek to impress their own nature on a mixed entity in terms of: 1) character (particle-like, wave-like); 2) location (in space or in time); and 3) temporal advance (rapid or slow/stationary). KE has its chance to prevail under two conditions: 1) when rest mass is very tiny, and 2) when rest mass velocity approaches the speed of light relative to some observer, thereby dwarfing the rest mass with the KE of motion. This results in space contraction and time dilation of the (mixed entity) mass object as measured from said observer's inertial system.[9]

**Collapse**: When work is done on rest mass, KE is created. The potential mass of this KE is in space where it constitutes a wave packet accompanying the rest mass as it moves. The rest mass may give up (transfer) this energy if it encounters a barrier such as a measuring instrument. This transfer of time-residing KE collapses the wave packet in space.

The collapse of a de Broglie wave packet is essentially the same as the collapse of waveform photon (probabilistic) potential mass. In both cases there is KE in time whose oscillation creates dependent space-progressing, waveform potential mass via $E = mc^2$. Collapse of these waves is nonlocal because they owe their occurring space presence to KE residing in orthogonal time. The collapse of the mathematical wave function $\Psi$ mirrors the collapse of packetized de Broglie waves.

---

[9] See the author's essay *"Minkowski and Special Relativity: Does His Spacetime Geometry Explain Space Contraction?"*: https://arxiv.org/abs/1602.02829

## 12.0   The Measurement Problem

Erwin Schrödinger assumed that the electron itself was a wave that his equation described; others took the electron to be a space-discrete particle and had to connect it and the wave equation via probability. In reality his equation describes potential mass: the space-progressing, waveform, probabilistic stored mass of the KE joined to electron rest mass. To repeat: the Schrödinger wave equation is modeling KE's potential mass which has the waveform and not the electron's rest mass which has the particle-form. Of course, they are inseparable

The idea that the wave function must describe the electron's rest mass follows from the (classical) assumption that both KE and the (potential) mass it stores are formless quantities with no presence in a dimension. Attributing wave character to the rest mass particle made the latter appear probabilistic in its existence, in its state or in its space location.

From this disconnect endless difficulties of interpretation have arisen over the years. Known collectively as "the measurement problem," these difficulties have led to numerous questions. Why is collapse necessary and what does it mean? Why only probability values from our equation? If rest mass (the electron) can be smeared over space, how does one get from there to matter being discrete? How can one get real knowledge of the quantum world using instruments obeying classical physics?

The "solutions" of the measurement problem over the decades extrapolate from our common experiences rather than looking at ontology. Working physicists ignore the problem entirely ("shut up and calculate").

Photon and electron wave collapse have already been explained as a consequence of occurring (waveform) potential mass in space being dependent upon an occurring entity in time. This is the case where something widespread in space (potential mass waves) has a single point of failure in the time dimension. Remaining issues of the measurement problem can be addressed by looking at the theory of particle superposition and the famous thought experiment it spawned.

## 13.0   Schrödinger's Cat

Explaining how the discrete electron can act like a continuous wave has led to the theory of **superposition**. Waveform superposition is something well-known and universally accepted both physically and mathematically. Waves can overlap in space and reinforce (or diminish) each other; energy oscillations (spin) can overlap and be "undefined." But applying superposition to static material entities means an existing rest mass can be in two places or states at once. This "solution" got a bit of mild mockery from Erwin Schrödinger when he published his famous thought experiment involving a cat that was both dead and alive. Schrödinger's thought experiment projects the probabilistic state of an unstable atom on to a cat and the details are familiar.

Assume there is a heavy atom (many electrons/protons) with an alpha particle oscillating within the potential well of this atom's nucleus. Like the much lighter electron, the energetic alpha particle (two protons and two neutrons bound together by the strong force) has KE joined to its rest mass. The KE entity has its potential (stored) mass accompanying the alpha particle rest mass as a standing wave that can be modeled by the wave function $\Psi$.

The wave function for the confined alpha particle yields a smeared probability density field that corresponds to the (likely) position of the oscillating particle. A portion of this field will extend beyond the potential barrier limits (quantum tunneling [16]). From this a probable particle release (decay) rate per hour or per day may be calculated. If the alpha particle remains in the nucleus the cat lives; if the alpha particle escapes the nucleus the lethal causal chain (detector, hammer, poison) is triggered and the cat dies. The wave function shows both cases (solutions) simultaneously and so the inference is that the cat is both dead and alive; these states are superposed.

The cat's state depends on the location of the alpha particle's rest mass: either inside the nucleus, alive, or outside the nucleus, dead. But the wave function is only characterizing the stored (potential) mass of the particle's KE, the latter having joined with the alpha particle's rest mass via work done. While a portion of the standing wave of potential mass may be outside the potential barrier of the nucleus, said potential mass waveform is merely (collapsible) objective probability. This potential mass waveform has no connection to, or effect upon, the causal chain that kills the cat; it is the rest mass that has that connection. Put another way, the wave function does not apply to (does not model) that entity (the alpha particle's rest mass) that can lethally affect the cat.

Arguing that superposition allows a rest mass particle to have two space locations at the same time is a consequence of not understanding what the wave function is modelling. Superposition applies to waveforms, in this case to space-continuous (probabilistic) potential mass as a standing wave. Superposition can also apply to spin orientation (up or down) because spin (angular momentum) is an expression of oscillatory KE. Superposition does not apply to material, particle-form entities such as a cat or an alpha particle's rest mass. The cat is never in a mixed, dead/alive, state. Curiosity kills cats, not probability waves.

## 14.0    Reductionism

Physicist John Archibald Wheeler writes, "There is not a single sight, not a single sound, not a single sense impression which does not derive in the last analysis from one or more elementary quantum phenomena [8, p.9]." Nobel laureate Steven Weinberg agrees, "Physicists and their apparatus must be governed by the same quantum mechanical rules that govern everything else in the universe. But these rules are expressed in terms of a wavefunction… [17]."

This is **ontological reductionism**: the argument that entities at a certain level can only be understood as collections or combinations of simpler entities at lower levels. The "ultimate laws of nature" therefore operate at the very bottom of the hierarchy of being.

Ontological reductionism has its place in scientific inquiry but it needs to be balanced by recognition that properties can emerge from the whole and not from the parts. Temperature characterizes a gas but not individual gas molecules. Saltiness characterizes sodium chloride, but not sodium nor chlorine. "Orbiting" electrons and their vibrating atoms/molecules are as much occurrence as existence. But several levels up this activity is the basis of the existing, bulk properties of solids, liquids or gases. What is discrete, oscillating, and KE at the lowest level becomes continuous, existing and potential energy only a few levels up. Those who deny emergent properties and insist that the wave function applies to human scale material objects are making an expression of faith that is based on an uncritical acceptance of reductionism.

## 15.0    Realism in Physics

Niels Bohr counseled a generation or two of physicists to refrain from speculating about quantum causes. But the human mind is not built that way; humans always want an explanation. But explaining quantum mysteries and paradoxes within the existence, mass and space framework is not possible and this has led to speculation on fanciful, made-up items such as supersymmetry, braneworlds or a multiverse.  Jim Baggott calls this "fairy tale physics [18, p.286]." All ages have their silly ideas; ours may be unique in the brain power (or at least the training) of the advocates. As an antidote, it is best to stick with a reality that is based simply on the concept of entity plus what can be measured: mass, energy, space, time. In this view, physical reality consists of entities that can store each other and either exist or occur alone (pure entities) or combine and do both simultaneously (mixed entities).

## 15.1 Different Entity Types have a Different Physics

Physics studies entities of three types. Not surprisingly, each type has its own characteristic form: 1) particle-form; 2) waveform; and 3) mixed-form. In broad outline, each type has its own physics.

Pure rest mass entities (inertial matter, no KE) have the particle-form and obey the space-stationary side of classical physics. Pure KE entities (EM radiation quanta) have the waveform and obey classical optics and electrodynamics (Maxwell's equations).
There are two areas where classical (pure entity) physics does not apply. First, classical physics does not apply to mixed-form entities where KE combines with rest mass and the KE is significant relative to the rest mass (i.e., if the rest mass is tiny or the velocity is extreme). Hence for the microworld quantum mechanics and the wave function $\Psi$ are necessary. Second, classical physics does not apply where quanta are unstable or cross the existence-occurrence, mass-energy divide. This would include pair production (a photon becoming an electron-positron pair), particle annihilation (electron meets positron) and particle disintegration (a muon).

Classical physics is perfectly valid for stationary material objects except at the quantal level where energy and matter waves become factors. Thus, bulk material objects, regarded as media, obey familiar-classical equations for statics (distribution of forces), stress/strain (Hooke's law) and hosted waves (sound waves, water waves, etc.). If the medium itself is uniform then the equations are straightforward; otherwise, they are the sum of local calculations. Classical mechanics (moving bodies) is very good (not perfect) for terrestrial matter-in-motion providing velocity (and hence KE) is small and rest mass is large.

Classical optics and electrodynamics are perfectly valid for radiation except when photons interact with sub-atomic particles since this invokes mixed-entity behavior requiring quantum physics. Such quantal interactions involve charges (and their tiny masses) interacting at small distances and low field strengths making QED necessary. This same limitation applies to condensed matter physics.

<u>In general</u>, classical physics applies to pure form entities. It is wave mechanics that absolutely depends upon a lack of purity of form. Wave mechanics cannot be used for pure form entities, namely in-flight EM radiation (Maxwell equations) or a boulder with zero momentum (statics, stress/strain). Wave mechanics and Heisenberg uncertainty require the joining of matter (rest mass) and energy; even that joining must have rest mass reduced to the vanishing point.

The question as to what separates classical physics from quantum physics has been a source of controversy for about a century. The temptation to regard wave mechanics as foundational (the "queen of physics") has proved irresistible for most; history is full of similar pronouncements. Despite the assertions of Wheeler and Weinberg, our laboratory instruments are not governed by quantum mechanical rules nor are they subject to the Heisenberg uncertainty relation. The electrons in our instruments may be in constant motion and describable by the wave function, but several levels up this electron KE has become binding energy in a static crystalline structure which exists.
Quantum physics is very different from classical physics and for good reason: they describe different entity configurations. There is a defined, ontological separation between the microworld and the macroworld; it is based not on size, but on pure entities versus mixed entities. The tradition is to view KE as an inert quantity. In reality KE is an entity with features of its own. When joined to rest mass these features compromise what we expect from stable rest mass. The contradictory features of rest mass joined to KE have led many to confidently proclaim the death of realism; they do this without even having a plausible theory of what the wave function represents.

In a letter to Otto Stern, Einstein wrote: "I've thought a hundred times more about quantum problems than about the general relativity theory." And to Max Born about the quantum theory he wrote: "I hope that someone will discover a more realistic way, or rather a more tangible basis than it has been my lot to find."

Einstein was at heart a 19th century realist who didn't like randomness; this despite the fact that he compromised radiation's continuity by quantizing it, thus making its interaction discrete. Perhaps he would accept that quantized, time-residing radiation energy interacting with orthogonal space-residing matter could not be causal/continuous and must instead yield random events. He also argued that quantum mechanics is incomplete. He was correct in that current QM is based on a flawed (incomplete) ontology.

**16.0    Conclusions:**

The photon is not unitary. It is an entity with two identities, energy residing in time and $E = mc^2$-derived probability waves advancing in space. Failure to recognized this has been a major obstacle to understanding radiation.

- **Radiation is the realm of occurrence, energy and time; it has been forced to fit in with the classical realm of existence, mass and space.**

- **Forcing radiation into particle physics (mechanics) has left us with many insoluble issues: duality (the double slit and the MZI), the constant speed of light, uncertainty, collapse, nonlocality and the measurement problem (Schrödinger's cat).**

Reality consists of entities of mass and energy which reside in a dimension and may also join one with another:

- **Mass entities residing in space may join together (composite material objects, Cooper pairs).**

- **Energy entities residing in time may join together (entangled photons).**

- **And the energy entity may join with the mass entity (matter-in-motion).**

Reality has a certain subtlety and symmetry in terms of equivalent constituents (mass and energy) residing in equivalent, hosting dimensions (space and time), all accessible via events that feature rest mass, kinetic energy and discrete location in space and time.

\* \* \* \* \* \* \* \* \* \*

# REFERENCES

[1] Klevgard, P.A., "Is the photon really a particle?" Optik, Vol. 237, 2021, https://doi.org/10.1016/j.ijleo.2021.166679.

[2] Koks, D., "What is Relativistic Mass?" http://math.ucr.edu/home/baez/physics/Relativity/SR/mass.html (10 December 2019).

[3] Shimony, A., "The Reality of the Quantum World," Scientific American, v258 n1 (Jan. 1988).

[4] Pais, A., [Niels Bohr's Times], Oxford Univ. Press (1991).

[5] Dongen, J. van, "The interpretation of the Einstein-Rupp experiments and their influence on the history of quantum mechanics," arXiv:physics/0709.3226 (3 January, 2020).

[6] "Elitzur–Vaidman bomb tester," https://en.wikipedia.org/wiki/Elitzur%E2%80%93Vaidman_bomb_tester. (4 February 2020).

[7] Wheeler, J.A., "The 'past' and the 'delayed-choice' double-slit experiment," Mathematical Foundations of Quantum Theory, ed. A.R. Marlow, Academic Press, New York, 9–48 (1978). For a summary see https://en.wikipedia.org/wiki/Wheeler%27s_delayed_choice_experiment#Cosmic_interferometer (4 May 2020).

[8] Wheeler, J., A., "Hermann Weyl and the Unity of Knowledge, http://www.weylmann.com/wheeler.pdf (18, February 2020).

[9] Roussel, P., Stefan, J. "Is the Interpretation of Delayed-Choice Experiments Misleading?" arXiv:physics/0706.2596. (6 January 2020).

[10] Pais, A., [Inward Bound], Oxford Univ. Press, New York, (1986).

[11] MacKinnon, E., "De Broglie's thesis: A critical retrospective," American Journal of Physics, 44 (1976).

[12] MacKinnon; also Roberto de Andrade Martins, "Louis de Broglie's Struggle with the Wave-Particle Dualism, 1923-1925," http://quantum-history.mpiwg-berlin.mpg.de/eLibrary/hq1_talks/waveMech/23_martins (14 May 2020).

[13] Ford, P.J., Saunders, G.A., [The Rise of the Supercomputers], CRC Press (2004). Also at: http://qudev.ethz.ch/content/courses/phys4/studentspresentations/supercond/Ford_The_rise_of_SC_6_7.pdf (10 August, 2020)

[14] Das, A., et.al., "Entangling electrons by splitting Cooper pairs: Two-particle conductance resonance and time coincidence measurements," https://arxiv.org/pdf/1205.2455.pdf (5 September 2020)

[15] Humphreys, P., et al., "Deterministic delivery of remote entanglement on a quantum network." Nature Letters 13 June 2018, https://www.nature.com/articles/s41586-018-0200-5 Also: https://www.osa-opn.org/home/newsroom/2018/june/entanglement_on_demand/

[16] See http://hyperphysics.phy-astr.gsu.edu/hbase/quantum/barr.html#c1 (5 June 2020)

[17] Weinberg, S., "Einstein's Mistakes," Physics Today, v58, Issue 11, section 'Contra Quantum Mechanics,' http://physicstoday.scitation.org/doi/full/10.1063/1.2155755 (3 April 2020)

[18] Baggott, J., [Farewell to Reality: How Modern Physics Has Betrayed the Search for Scientific Truth], Pegasus Books, New York, (2013)

[19] Planck, M., Nobel Lecture: "The Genesis and Present State of Development of the Quantum Theory," https://www.nobelprize.org/prizes/physics/1918/planck/lecture/ (7 March 2020).